\documentclass[a4paper,conference]{IEEEtran}

\usepackage{etex}

\usepackage{graphicx,subfigure}
\usepackage{amsmath,amssymb,amsthm}
\usepackage[noadjust]{cite}
\usepackage{tikz}
\usetikzlibrary{positioning}

\allowdisplaybreaks

\newtheorem{lem}{Lemma}
\newtheorem{cor}{Corollary}

\newtheorem{prop}{Proposition}

\theoremstyle{definition}
\newtheorem{definition}{Definition}

\newtheorem{example}{Example}

\newcommand{\Ggraph}{\mathcal{G}}
\newcommand{\Lumping}{(\Pvec,g)}
\newcommand{\mblocked}[1]{#1^{(K)}}
\newcommand{\Set}[1]{\{#1\}}
\newcommand{\ent}[1]{H(#1)}
\newcommand{\binent}[1]{H_2(#1)}

\newcommand{\entrate}[1]{\bar{H}(\mathbf{#1})}

\newcommand{\loss}[2][\empty]{\ifthenelse{\equal{#1}{\empty}}{L(#2)}{L_{#1}(#2)}}
\newcommand{\lossrate}[2][\empty]{\ifthenelse{\equal{#1}{\empty}}{L(\mathbf{#2})}{L_{\mathbf{#1}}(\mathbf{#2})}}

\newcommand{\relLoss}[2][\empty]{\ifthenelse{\equal{#1}{\empty}}{l(#2)}{l_{#1}(#2)}}

\newcommand{\dom}[1]{\mathcal{#1}}

\newcommand{\Prob}[1]{\mathrm{Pr}(#1)}

\newcommand{\card}[1]{\mathrm{card}(#1)}

\newcommand{\Xvec}{\mathbf{X}}

\newcommand{\Yvec}{\mathbf{Y}}

\newcommand{\Zvec}{\mathbf{Z}}

\newcommand{\Pvec}{\mathbf{P}}
\newcommand{\muvec}{\boldsymbol{\mu}}

\newcommand{\Wvec}{\mathbf{W}}

\newcommand{\Amat}{\mathbf{A}}

\newcommand{\preim}[1]{g^{-1}(#1)}

\renewcommand{\card}[1]{\left|#1\right|}

\title{Graph-Based Lossless Markov Lumpings}

\author{
\IEEEauthorblockN{
  Bernhard C. Geiger\IEEEauthorrefmark{1}
, Christoph Hofer-Temmel\IEEEauthorrefmark{2}
}
\IEEEauthorblockA{
\IEEEauthorrefmark{1}
Institute for Communications Engineering, TU M\"unchen, Germany\\
\IEEEauthorrefmark{2}
Faculteit Militaire Wetenschappen, Nederlandse Defensie Academie, The Netherlands\\
geiger@ieee.org, math@temmel.me}
}

\newcommand{\Abstract}{
\begin{abstract}
We use results from zero-error information theory to determine the set of non-injective functions through which a Markov chain can be projected without losing information. These \emph{lumping functions} can be found by clique partitioning of a graph related to the Markov chain. Lossless lumping is made possible by exploiting the (sufficiently sparse) temporal structure of the Markov chain. Eliminating edges in the transition graph of the Markov chain trades the required output alphabet size versus information loss, for which we present bounds.
\end{abstract}
}
\newcommand{\Acknowledgements}{
\section*{Acknowledgments}
The authors thank Ali Amjad, Andrei Nedelcu, and Wolfgang Utschick for fruitful discussions. The work of Bernhard C. Geiger was partially funded by the Erwin Schr\"odinger Fellowship J 3765 of the Austrian Science Fund.
}
\newcommand{\Bibliography}{
\bibliographystyle{IEEEtran}
\bibliography{IEEEabrv,references}
}

\begin{document}
\maketitle
\Abstract{}
\section{Introduction}

\begin{figure}[t]
\centering
  \begin{tikzpicture}[-latex ,auto ,node distance =1.5 cm and 2.5 cm ,on grid ,thick ,state/.style ={ circle ,top color =white , bottom color = black!20 ,draw,black , text=black , minimum width =0.5 cm},rounded corners]
\tikzset{>=latex}
\node[state] (A) {$1$};
\node[state] (B) [below =of A] {$2$};
\node[state] (C) [left =of B] {$3$};
\node[state] (D) [above =of C] {$4$};

\path (A) edge [->] (C);
\path (D) edge [->] (A);
\path (C) edge [->] (B);
\path (B) edge [->] (D);
\path (D) edge [loop left] (D);
\path (C) edge [loop left] (C);
\path (B) edge [loop right] (B);
\path (A) edge [loop right] (A);

\draw[draw=red] (-3.5,-2) rectangle (-2,0.5); \draw[-] (-3.25,0.75) node {\textcolor{red}{$2'$}};
\draw[draw=red] (-0.5,-2) rectangle (1,0.5); \draw[-] (0.75,0.75) node {\textcolor{red}{$1'$}};
\end{tikzpicture}
\caption{The transition graph of an irreducible, aperiodic Markov chain with alphabet $\dom{X}=\{1,2,3,4\}$. The partition indicated by the red boxes induces a lumping function $g$, with $g(1)=g(2)=1'$ and $g(3)=g(4)=2'$. While $g$ is not invertible, side information about the previous state allows to determine the current state given only the lumped state: If the previous state is 1 and the current lumped state is 2' (box on the left), only state 3 is realizable.}
\label{fig:states}
\end{figure}
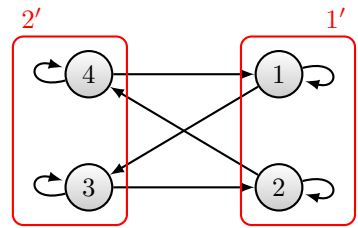

Large Markov models, common in many scientific disciplines, present a challenge for analysis, model parameter learning, and simulation: Language $n$-gram models~\cite[Ch.~6]{Manning_NLP} and models in computational chemistry and systems biology~\cite{Wilkinson_SystemsBiology}, for example, belong to this category. For these models, efficient simulation methods are as important as ways to represent the model with less parameters. A popular approach for the latter is lumping, i.e., replacing the alphabet of the Markov chain by a smaller one via partitioning. This partition induces a non-injective lumping function from the large to the small alphabet. While, in general, the lumped process has a lower entropy rate than the original chain, in~\cite{Geiger_Temmel__kLump} we presented conditions for lossless lumpings, i.e., where the original Markov chain and the lumped process have equal entropy rates. Specifically, the \emph{single entry} property we define in~\cite[Def.~3]{Geiger_Temmel__kLump} holds if, given the previous state of the Markov chain, in the preimage of the current lumped state only a single state is realizable, i.e., has positive probability (see Fig.~\ref{fig:states}).

The emphasis on whether a state is realizable, rather than on its probability, is also common in zero-error information theory. Typical problems in zero-error information theory are error-free communication~\cite{Shannon_ZeroError} (rather than communication with small error probabilities) and lossless source coding with side information~\cite{Witsenhausen_Confusability}. Both problems admit elegant graph-theoretic approaches which we recapitulate in Section~\ref{sec:prel}.

% Complementing his earlier results on communication with sufficiently small error probabilities, Shannon introduced zero-error information theory requiring error-free communication~\cite{Shannon_ZeroError}. Later, Witsenhausen~\cite{Witsenhausen_Confusability} defined zero-error source coding with side information, i.e., the scenario in which the receiver can decode the source signal without error, given some side information. Both problems, i.e., that of zero-error communication and that of zero-error source coding, have elegant graph-theoretic descriptions which we recapitulate in Section~\ref{sec:prel}.

In Section~\ref{sec:graphcomp}, we use these graph-theoretic approaches to find lossless lumpings for a given Markov chain. While the current state of the Markov chain cannot be inferred from its lumped image only, we require that it can be reconstructed by using the previous state of the Markov chain as \emph{side information} (cf.~Fig.~\ref{fig:states}). The lumpings fulfilling this requirement correspond to the possible clique partitions of a graph derived from the Markov chain. The method is universal in the sense that it only depends on the presence, but not the precise magnitude, of state transitions of the Markov chain. In Section~\ref{sec:lossycomp}, we relax the problem and reduce the output alphabet size of the lumping function by accepting that the lumped process has an entropy rate smaller than the original chain. We furthermore present bounds on the difference between these entropy rates.

By design, lossless lumpings are not efficient source codes. Thus, it cannot be assumed that the reduced output alphabet size is related to the Markov chain's entropy rate. Nevertheless, in Section~\ref{sec:block_source}, we evaluate our lossless lumping method from a source coding perspective by applying it to length-$K$ sequences of the original Markov chain. We show that the required size of the output alphabet never exceeds (and asymptotically approaches) the number of realizable length-$K$ sequences. Our lossless lumping method is thus an asymptotically optimal fixed-length, lossless source code.

Future work shall apply the presented lumping methods to practical examples from, e.g., chemical reaction networks or natural language processing. Furthermore, while the connection between lossless lumpings and zero-error information theory is interesting and revealing, searching lossless lumping functions via clique partitioning can be computationally expensive. We have reasons to believe that the search for lumping functions can be cast as a constrained optimization problem whose properties are currently under investigation. Finally, we believe that the results presented in this work can contribute to zero-error source coding of processes with memory, complementing available results on zero-error coding for channels with memory (see~\cite{Cohen_Fachini_Koerner__ZeroErrorCapacityOfBinaryChannelsWithMemory__IEEE_TIT_2016} and the references therein). 
% The extended version~\cite{Geiger_HoferTemmel__ZEGraph__arXiv} of our manuscript hints at first results.
Section~\ref{sec:discussion} hints at first results.

\section{Preliminaries from Zero-Error Information Theory}
\label{sec:prel}
Throughout this work, $\log$ denotes the natural logarithm, i.e., entropies and entropy rates are measured in nats.

Let $\Xvec:=(X_n)_{n=1}^\infty$ be an irreducible, aperiodic, stationary Markov chain with finite alphabet $\dom{X}:=\{1,\dots,N\}$, transition probability matrix $\Pvec$, and invariant distribution vector $\muvec$. The adjacency matrix $\Amat$ is defined by $A_{x,x'}:=\lceil P_{x,x'}\rceil$. We say a state $x$ can \emph{access} another state $x'$, if $P_{x,x'}>0$ ($A_{x,x'}=1$). We abbreviate $X_m^n:=\{X_m,X_{m+1},\dots,X_n\}$. The $K$-fold blocked process $\mblocked{\Xvec}$ given by $\mblocked{X}_n := X_{(n-1)K+1}^{nK}$ is also Markov. Every length-$K$ sequence of $\Xvec$ is a state of $\mblocked{\Xvec}$.

Let $\Ggraph:=(\dom{X},E)$ be a graph with vertices $\dom{X}$ and edges $E\subseteq[\dom{X}]^2$, where $[\dom{X}]^2$ is the set of two-element subsets of $\dom{X}$. A set $S\subseteq \dom{X}$ is a clique, if $[S]^2 \subseteq E$, and an independent set, if $[S]^2 \cap E=\emptyset$. The clique number $w(\Ggraph)$ and independence number $\alpha(\Ggraph)$ are the size of $\Ggraph$'s largest clique and independent set respectively. A clique partition of $\Ggraph$ is a partition of $\dom{X}$ into cliques of $\Ggraph$. The clique partition number $\gamma(\Ggraph)$ is the size of the smallest clique partition of $\Ggraph$. The chromatic number $\chi(\Ggraph)$ is the minimum number of colours needed to paint $\dom{X}$ without having same-coloured neighbours.

The complement graph $\overline\Ggraph$ has vertex set $\dom{X}$ and edge set $[\dom{X}]^2\setminus E$. Edge-duality identifies cliques of $\Ggraph$ and independent sets of $\overline\Ggraph$ and vice-versa, whence $w(\Ggraph)=\alpha(\overline\Ggraph)$ and $\gamma(\Ggraph)=\chi(\overline\Ggraph)$. For further details on graph theory see~\cite{Diestel__GraphTheory__Springer_2005}.

Let $\dom{Y}:=\{1,\dots,M\}$. We consider a discrete, memoryless channel (DMC) with input alphabet $\dom{X}$ and output alphabet $\dom{Y}$, defined by the transition probability matrix $\mathbf{W}$, where $W_{x,y} := \Prob{Y_n=y|X_n=x}$. In the case of a deterministic channel, i.e., one in which $W_{x,y}\in\{0,1\}$ and $M\le N$, we can describe the channel by a \emph{lumping function} $g{:}\ \dom{X}\to\dom{Y}$ and call $\Yvec$ defined by $Y_n:=g(X_n)$ the lumped process.

\begin{definition}\label{def:channelgraph}
Let $\Ggraph_W:=(\dom{X},E_W)$ be the \emph{(channel) confusion graph}, where
\begin{align}
 \{x_1,x_2\}\in E_W
 \Leftrightarrow
 \exists y\in\dom{Y}{:}\ \lceil W_{x_1,y}\rceil\cdot\lceil W_{x_2,y}\rceil = 1
 \,.
\end{align}
In the case of a deterministic channel, i.e., a lumping, denote the confusion graph by $\Ggraph_g:=(\dom{X},E_g)$.
\end{definition}
The confusion graph connects two vertices if the channel confuses them with positive probability, i.e., if there exists at least one element in the output alphabet to which both inputs can be mapped. If the channel is deterministic, then the confusion graph $\Ggraph_g$ has a simple structure.
\begin{lem}\label{lem:functgraph}
The confusion graph $\Ggraph_g$ consists of isolated cliques induced by the preimages of the lumping function $g$. Hence,
\begin{equation}
 E_g = \bigcup_{y\in\dom{Y}} \left[\preim{y}  \right]^2
 \,.
\end{equation}
\end{lem}

The confusion graph is exactly the graph used in Shannon's original paper~\cite{Shannon_ZeroError} and the complement of the graph in~\cite[Sec.~III]{Korner_ZeroError}. The confusion graph determines the zero-error capacity $C(W)$ of the channel. The number of messages that can be transmitted reliably via one channel use is the independence number of its confusion graph $\alpha(\Ggraph_W)$. For $K$ channel uses, one requires the $K$-fold normal product of $\Ggraph_W$ with itself: $\Ggraph_W^{\wedge K}:=(\dom{X}^K,E_W^{\wedge K})$, where $\{x_1^K,{x'}_1^K\}\in E_W^{\wedge K}$, if $\{x_i,x'_i\}\in \dom{X}\cup{}E_W$ for all $i=1,\dots,K$. In the limit, one has the zero-error capacity $C(W):=\sup_K 1/K \log\alpha(\Ggraph_W^{\wedge K})\ge \alpha(\Ggraph_W)$. In the case of a deterministic channel, the number of messages that can be transmitted reliably in one channel use is $\alpha(\Ggraph_g)=\gamma(\Ggraph_g)=\card{\dom{Y}}=M$, where $\card{A}$ is the cardinality of the set $A$. Since the normal product of a graph of isolated cliques is again a graph of isolated cliques, one has $C(g)=\log M$, cf.~\cite[p.~2209]{Korner_ZeroError}. For such channels, separating source and channel coding is optimal~\cite[Prop.~1]{Nayak_ZESCCoding}.

Let $X$ and $Z$ be two RVs with a joint distribution having support $\dom{S}_1:=\{(x,z)\in\dom{X}\times\dom{Z}{:}\ \Prob{X=x,Z=z}>0\}$.
\begin{definition}\label{def:generalchargraph}
Let $\Ggraph_{(X,Z)}:=(\dom{X},E_{(X,Z)})$ be the \emph{characteristic graph} of $(X,Z)$, where $\{x,x'\}\in E_{(X,Z)}$, if
\begin{equation}\label{eq:generalchargraph}
 \forall z\in\dom{Z}{:}\ \Prob{X=x,Z=z}\Prob{X=x',Z=z}=0
 \,,
\end{equation}
i.e., if there is no $z$ such that $(x,z)\in\dom{S}_1$ and $(x',z)\in\dom{S}_1$.
\end{definition}

In other words, the characteristic graph connects two vertices, if the \emph{side information} $Z$ distinguishes between them. The characteristic graph is the complement of the graph defined by Witsenhausen~\cite{Witsenhausen_Confusability}. It determines the smallest number of messages that the transmitter must send to the receiver, such that the latter can reconstruct $X$ with the help of the side information $Z$. For a single transmission, the required number of messages is the clique partition number $\gamma(\Ggraph_{(X,Z)})$. For $K$ independent instances of $(X,Z)$, one requires the $K$-fold co-normal product of $\Ggraph_{(X,Z)}$ with itself: $\Ggraph_{(X,Z)}^{\vee K}:=(\dom{X}^K,E_{(X,Z)}^{\vee K})$, where $\{x_1^K,{x'}_1^K\}\in E_{(X,Z)}^{\vee K}$, if $\{x_i,x'_i\}\in E_{(X,Z)}$ for at least one $i=1,\dots,K$. In particular, $\Ggraph_{(X,Z)}^{\vee K}=\Ggraph_{\mblocked{(X,Z)}}$, the characteristic graph of the $K$-fold blocked process. The number of bits required to convey $K$ instances is thus $\log \gamma(\Ggraph_{(X,Z)}^{\vee K})\le K \log\gamma(\Ggraph_{(X,Z)})$.

The characteristic graph $\Ggraph_{(X,Z)}$ depends only on the source and connects messages $X$ that the channel may confuse, given the receiver has side information $Z$. The confusion graph $\Ggraph_W$ depends only on the channel and connects messages that the channel confuses. If the edge set of the latter is a subset of the edge set of the former, the channel confuses only messages that can be distinguished by incorporating the side information. This is the statement of
\begin{prop}\label{prop:zerocomm}
 $E_W\subseteq E_{(X,Z)} \Leftrightarrow \ent{X|Y,Z}=0$.
\end{prop}
Proposition~\ref{prop:zerocomm}, proved in Section~\ref{proof:zerocomm}, generalizes easily to multiple channel uses by considering the corresponding graph products.

\section{Graph-Based Lossless Markov Lumpings}
\label{sec:graphcomp}

We use results from zero-error information theory to construct a lumping of a Markov chain such that the original Markov chain can be recovered without error. To this end, we assume that, for the reconstruction of $X_n$, the receiver has the previous state $X_{n-1}$ as side information. This \emph{temporal side information} determines the characteristic graph. A clique partition of this graph defines a lumping function $g$, whose confusion graph (Definition~\ref{def:channelgraph}) is a subset of the Markov chain's characteristic graph. Then, Proposition~\ref{prop:zerocomm} guarantees that the original chain can be perfectly reconstructed from its initial state and the lumped process.
The remainder of this section makes these statements precise.

\begin{definition}\label{def:chargraph}
Let $\Ggraph_\Xvec:=(\dom{X},E_\Xvec)$ be the \emph{characteristic graph} of $\Xvec$, where
\begin{align}
 \{x_1,x_2\}\in E_\Xvec
 \Leftrightarrow
 \forall x\in\dom{X}{:}\ A_{x,x_1}A_{x,x_2} =0
 \,.
\end{align}
\end{definition}

In other words, the characteristic graph of a Markov chain connects two states, if every state can only access one of them. Since the Markov chains considered in this work are irreducible, the invariant distribution vector is positive and Definition~\ref{def:chargraph} coincides with Definition~\ref{def:generalchargraph} for a source $X_n$ with side information $X_{n-1}$.

\begin{example}
Consider the Markov chain in Fig.~\ref{fig:states}. Its characteristic graph has edge set $E_\Xvec=\Set{\Set{1,2},\Set{3,4}}$. Both edges are cliques, and together they partition $\dom{X}$.
\end{example}

Choose an arbitrary clique partition of $\Ggraph_\Xvec$, enumerate the cliques, and define $g$ such that it maps each vertex in $\dom{X}$ to the index of its containing clique. This way, $g$ assigns different values to vertices within different cliques. According to Lemma~\ref{lem:functgraph}, the confusion graph $\Ggraph_g$ of $g$ consists exactly of the cliques of the chosen clique partition of $\Ggraph_\Xvec$, only that these cliques are isolated in $\Ggraph_g$. This ensures that $E_g\subseteq E_\Xvec$.
Let $Y_n:=g(X_n)$ define the lumped process $\Yvec$.
Hence, by Proposition~\ref{prop:zerocomm}, we have
\begin{equation}\label{eq:upperboundzero}
 \ent{X_n|Y_n,X_{n-1}}=0
 \,.
\end{equation}
Let $\entrate{\Xvec}$ and $\entrate{\Yvec}$ be the entropy rates of $\Xvec$ and $\Yvec$ respectively.
It is easy to see that the tuple $\Lumping$ fulfils the single-entry property~\cite[Def.~10]{Geiger_Temmel__kLump}.
Thus, the lumping is lossless in the sense of a vanishing information loss rate, i.e.,
\begin{equation}
 \entrate{\Xvec|\Yvec}
 :=\lim_{n\to\infty} \frac{1}{n} \ent{X_1^n|Y_1^n}
 =\entrate{\Xvec}-\entrate{\Yvec}=0
 \,.
\end{equation}
This follows from the chain rule $(a)$, the fact that conditioning reduces entropy $(b)$, and stationarity of $\Xvec$ $(c)$:
\begin{subequations}\label{eq:upperboundproof}
 \begin{align}
 \entrate{\Xvec|\Yvec}
 &\stackrel{(a)}{=}\lim_{n\to\infty} \frac{1}{n}\sum_{i=1}^n\ent{X_i|Y_1^n,X_1^{i-1}}\\
 &\stackrel{(b)}{\leq} \lim_{n\to\infty} \frac{1}{n}\sum_{i=1}^n\ent{X_i|Y_i,X_{i-1}}\\
 &\stackrel{(c)}{=} \ent{X_2|Y_2,X_{1}}\,.
\end{align}
\end{subequations}
The last term vanishes because $g$ is such that~\eqref{eq:upperboundzero} holds for all $n$. With this we have proven

\begin{cor}\label{cor:zerocomm}
If, for a given Markov chain $\Xvec$, the lumping function $g$ satisfies $E_g\subseteq E_\Xvec$, then the lumping is lossless, i.e., $\entrate{\Xvec|\Yvec} =0$.
\end{cor}

Not only is the proposed lumping method lossless in the sense of Corollary~\ref{cor:zerocomm}, the original Markov chain can be perfectly reconstructed from its initial state $X_1$ and from the lumped process $\Yvec$. The initial state $X_1$ and the state $Y_2$ of the lumped process together determine the state $X_2$ of the original Markov chain. Then, $X_2$ acts as side information to reconstruct $X_3$ from $Y_3$, etc.

We investigate the size $M$ of the output alphabet required for $g$ to be lossless. An optimal lumping function $g$ induces the smallest possible partition of $\dom{X}$, i.e., $M=\gamma(\Ggraph_g)=\gamma(\Ggraph_\Xvec)$. From Definition~\ref{def:chargraph} follows that no two states accessible from a given state $x\in\dom{X}$ can be connected in $\Ggraph_\Xvec$. Hence, if $d_{\max}$ is the maximum out-degree of the transition graph associated with $\Pvec$, i.e.,
\begin{equation}
 d_{\max} := \max_x \sum_{x'\in\dom{X}} A_{x,x'}
\end{equation}
then $\Ggraph_\Xvec$ contains at least $d_{\max}$ cliques. We recover

\begin{prop}[{\cite[Prop.~3]{Geiger__Markov__arXiv}}]\label{prop:Mbound}
$M\ge d_{\max}$.
\end{prop}

Witsenhausen~\cite[Prop.~1]{Witsenhausen_Confusability} showed that this lower bound can be achieved using the side information, which is available at both ends. The achievable scheme requires that, for every state of the side information $X_{n-1}$, a separate lumping function is used. Our restriction to a single lumping function leads to an output alphabet size generally larger than $d_{\max}$. However, if $\Amat$ is sufficiently sparse, then the presence of side information at the receiver helps to make the output alphabet size still strictly smaller than $N$.

\begin{example}
Consider the Markov chain in Fig.~\ref{fig:states} and assume that all transitions have probability 0.5. By symmetry, it follows that $\ent{X_n}=\log N=\log 4$ and $\entrate{\Xvec}=\log M = \log 2$. The output alphabet size is optimal in terms of Proposition~\ref{prop:Mbound}: $\ent{Y_n}=\entrate{\Yvec}=\log M=\log d_{\max}=\log 2$.
\end{example}

The proposed lumping method depends only on the location of zeros in the adjacency matrix $\Amat$. It follows that the method is \emph{universal} in the sense that the obtained lumping function $g$ is lossless for every Markov chain with adjacency matrix $\Amat$. Moreover, $g$ is lossless for every stationary process, for which the non-zero one-step transition probabilities are modelled by $\Amat$. Equations~\eqref{eq:upperboundproof} do not require Markovity of $\Xvec$, whence Corollary~\ref{cor:zerocomm} remains valid. However, our lumping method is only useful for Markov chains (or stationary processes) with a \emph{deterministic temporal structure}, i.e., for sparse matrices $\Amat$.

\begin{example}
Suppose that $\Pvec$ is a positive matrix, collecting the conditional probability distribution of two consecutive samples of $\Xvec$. Hence $\Amat$ is a matrix of ones, and the edge set $E_\Xvec$ of the characteristic graph $\Ggraph_\Xvec$ is empty. Thus, $M=\gamma(\Ggraph_\Xvec)=N$. The only lossless lumping functions are permutations, hence lumping does not reduce the alphabet size.
\end{example}

Note finally that instead of defining $g$ via a clique partition of $\Ggraph_\Xvec$, one can also define a \emph{stochastic lumping $\Wvec$} via a clique covering of $\Ggraph_\Xvec$. This still ensures that $E_\Wvec\subseteq E_\Xvec$ holds and that the statement of Corollary~\ref{cor:zerocomm} remains valid. While clique covering leads to additional freedom in the design of the lumping, it does not reduce the required output alphabet size compared to clique partitioning: If two cliques $S_1$ and $S_2$ cover a subset of the vertices $\dom{X}$, then the two cliques $S_1$ and $S_2\setminus S_1$ partition it.

\section{Graph-Based Lossy Markov Lumpings}
\label{sec:lossycomp}

We generalize the characteristic graph of the Markov chain by eliminating edges from its transition graph (i.e., ones in its adjacency matrix $\Amat$) if the transition probabilities fall below a certain threshold:

\begin{definition}\label{def:epschar}
For $\varepsilon>0$, the \emph{$\varepsilon$-characteristic graph} of $\Xvec$ is the graph $\Ggraph_\varepsilon:=(\dom{X},E_\varepsilon)$, where
\begin{align}
 \{x_1,x_2\}\in E_\varepsilon
 \Leftrightarrow
 \forall x\in\dom{X}{:}\,
 \lceil P_{x,x_1}\!-\varepsilon\rceil\cdot\lceil P_{x,x_2}\!-\varepsilon\rceil
 = 0
 \,.
\end{align}
\end{definition}

Definition~\ref{def:epschar} is equivalent to Definition~\ref{def:chargraph}, if $\Amat$ is defined by $A_{x,x'}:=\lceil P_{x,x'}-\varepsilon\rceil$. Decreasing the number of ones in $\Amat$ can only increase the number of edges in the characteristic graph, which in turn can only make the cliques larger and the clique partition number smaller. Hence, $E_\Xvec\subseteq E_\varepsilon$ and $\gamma(\Ggraph_\Xvec)\ge\gamma(\Ggraph_\varepsilon)$. By eliminating edges, one may trade information loss for alphabet size. For the former, in Section~\ref{proof:epserr}, we prove a bound depending on $\varepsilon$, the number $N$ of vertices, and the cardinality of the output alphabet $M$:

\begin{prop}\label{prop:epserr}
Take $\varepsilon<1/N$ and $E_g\subseteq E_\varepsilon$, then
\begin{equation}\label{eq:epserr}
 \entrate{\Xvec|\Yvec}
 \le (N-M)\varepsilon\left(1-\log\varepsilon\right)
 \le N\binent{\varepsilon}
 \,,
\end{equation}
where $\binent{p}:= -p\log p-(1-p)\log(1-p)$. The first inequality already holds for $\varepsilon<1/e$.
\end{prop}

Applying Proposition~\ref{prop:epserr} to $\varepsilon=0$ recovers Corollary~\ref{cor:zerocomm}. The following example illustrates that if the entropy rate of $\Xvec$ falls below the bound in Proposition~\ref{prop:epserr}, the lumped process $\Yvec$ can become trivial.

\begin{example}
Suppose that
\begin{equation}
 \Pvec=\begin{pmatrix}
  1-\varepsilon
  & \varepsilon
  \\\varepsilon
  & 1-\varepsilon
 \end{pmatrix}
 \,.
\end{equation}
It follows that $\mu_1=\mu_2=1/2$ and that $\entrate{\Xvec}=\binent{\varepsilon}$. Moreover, as $\Ggraph_\varepsilon$ is fully connected, $g$ is constant with $M=1$. Thus, $\Yvec$ is a constant process and $\entrate{\Yvec}=0$.
\end{example}

Reconstructing $\Xvec$ from $\Yvec$ (with small probability of error) requires reconstruction methods more sophisticated than those for the lossless lumping method introduced in Section~\ref{sec:graphcomp}. Given knowledge of the previous state $X_{n-1}$ and the current lumped state $Y_n$, the current state $X_n$ can not be reconstructed without error. Hence, the side information used for reconstructing the next state might not be correct, which leads to error propagation.

\section{A Source Coding Perspective on Lossless Markov Lumpings}
\label{sec:block_source}

The intended application of the lumping method introduced in Section~\ref{sec:graphcomp} -- model reduction in speech/language processing~\cite{Manning_NLP} or systems biology~\cite{Wilkinson_SystemsBiology} -- imposes several restrictions.
The lumping is a time-invariant, preferably deterministic mapping from the large alphabet $\dom{X}$ to a smaller alphabet $\dom{Y}$ and operates on a symbol-by-symbol basis in order to represent a partition of the original alphabet. These restrictions -- \emph{stateless, fixed-length}, and \emph{symbol-by-symbol} -- make our proposed method an inefficient source code.
Despite this apparent incompatibility, we critically evaluate our lossless lumping method from a source coding perspective.

First, our lumping method can be used as a (universal) pre-processing step, after which more sophisticated compression schemes follow. For example, it can be easily extended to a variable-length symbol-by-symbol scheme by, e.g., optimal Huffman coding of the lumped states.

Second, we may still require the lumping to be stateless and fixed-length, but define the lumping function $g$ on the $K$-fold Cartesian product $\dom{X}^K$.
Hence, $g$ lumps sequences of length $K$ rather than states.
Due to the deterministic temporal structure of $\Xvec$, the alphabet size for lumping these length-$K$ sequences is not larger than the number of realizable sequences of this length.
In other words, our scheme is at least as good as, and asymptotically equivalent to, any fixed-length, lossless coding scheme that en-/decodes sequences independently of each other. To show this, let $\lambda$ be the largest eigenvalue of the adjacency matrix $\Amat$. If the Markov chain $\Xvec$ has adjacency matrix $\Amat$, the logarithm of $\lambda$ bounds the entropy rate of $\Xvec$ from above, i.e., $\entrate{\Xvec}\le \log \lambda$~\cite{Delvenne_Centrality,Burda_MERW}. In Section~\ref{proof:optimal}, we prove
\begin{prop}\label{prop:optimal}
For each $K$, let $g_K{:}\ \dom{X}^K\to \dom{Y}_K$ be the optimal lumping function for the Markov chain $\mblocked{\Xvec}$, i.e., it induces the smallest clique partition of its characteristic graph $\Ggraph_{\mblocked{\Xvec}}$. Let $M_K:=\card{\dom{Y}_K}$. Let the set of realizable states of $\mblocked{\Xvec}$ be
\begin{equation}
 \dom{S}_K := \{x\in\dom{X}^K{:}\ \Prob{\mblocked{X}_n=x}>0\}
 \,.
\end{equation}
Then, $M_K\le\card{\dom{S}_K}$ and
\begin{equation}\label{eq:optimal}
 \lim_{K\to\infty} \frac{\log M_K}{K} = \log\lambda
 \,.
\end{equation}
\end{prop}

\begin{example}
If $K=1$, then $M=M_1\le\card{\dom{S}_1}=\card{\dom{X}}=N$. If $K=2$, then $M_2\le \card{\dom{S}_2}=\sum_{i,j} A_{i,j} \le N^2$.
\end{example}

While $M_K\le\card{\dom{S}_K}$, especially for small $K$ and sparse $\Amat$, the inequality may be strict. This advantage disappears for increasing $K$ due to the Markov property, and the required alphabet size approaches the number of realizable length-$K$ sequences, which for large $K$ behaves like $\lambda^K$~\cite{Burda_MERW}. Thus, while our lossless lumping method is asymptotically optimal in the sense of Proposition~\ref{prop:optimal}, for the intended application of reducing the alphabet it seems to be most efficient when applied symbol-by-symbol.

\section{Proofs}
\subsection{Proof of Proposition~\ref{prop:zerocomm}}
\label{proof:zerocomm}

Let $Q_{x,z}:=\Prob{X=x,Z=z}$.
First, assume that $\ent{X|Y,Z}>0$.
There exist triples $(x,y,z)$ and $(x',y,z)$ such that
\begin{subequations}\label{eq:proof:prop1}
\begin{equation}
 \Prob{X=x,Y=y,Z=z}=Q_{x,z}W_{x,y}>0
\end{equation}
and
\begin{equation}
 \Prob{X=x',Y=y,Z=z}=Q_{x',z}W_{x',y}>0
 \,.
\end{equation}
\end{subequations}
Hence, each term of the products on right-hand sides above must be positive, from which $Q_{x,z}Q_{x',z}>0$ and $\lceil W_{x,y}W_{x',y}\rceil=1$ follows.
As a consequence, by the definitions of the channel confusion graph and the characteristic graph of $(X,Z)$, we have $\{x,x'\}\in E_W$ and $\{x,x'\}\notin E_{(X,Z)}$.
Thus, $E_W \not\subset E_{(X,Z)}$.

Second, assume that $E_W \not\subset E_{(X,Z)}$.
Then, there exists $\{x,x'\}\in[\dom{X}]^2$ such that $\{x,x'\}\in E_W$ and $\{x,x'\}\notin E_{(X,Z)}$.
It follows that there exists at least one $z'$ such that $Q_{x,z'}Q_{x',z'}>0$, and at least one $y'$ such that $W_{x,y'}W_{x',y'}>0$.
Hence, the two probabilities in equations~\eqref{eq:proof:prop1} are positive for $z=z'$ and $y=y'$.
Thus, $\ent{X|Y,Z}>0$.\qed

\subsection{Proof of Proposition~\ref{prop:epserr}}
\label{proof:epserr}

That $\entrate{\Xvec|\Yvec}\le\ent{X_2|Y_2,X_1}$ follows from~\eqref{eq:upperboundproof}. If we define $R_{x,y}:=\sum_{x'\in\preim{y}} P_{x,x'}$, then we get
\begin{equation}
 \ent{X_2|Y_2,X_1=x}
 = -\sum_{y\in\dom{Y}} \sum_{x'\in\preim{y}}
  P_{x,x'} \log\frac{P_{x,x'}}{R_{x,y}}
 \,.
\end{equation}
The assumption $E_g\subseteq E_\varepsilon$ implies that $g^{-1}(y)$ is a clique in $E_\varepsilon$, whence each $x\in\dom{X}$ can access at most one element in $g^{-1}(y)$ with a probability larger than $\varepsilon$. Hence, let $\hat{x}\in\preim{y}$ be such that for all other $x''\in\preim{y}\setminus\{\hat{x}\}$, $P_{x,x''}<\varepsilon$. Thus,
\begin{equation}\label{eq:boundR}
 R_{x,y}\le P_{x,\hat{x}}+\varepsilon \left(\card{\preim{y}}-1\right)
 \,.
\end{equation}
We derive the first inequality in~\eqref{eq:epserr}:
\begin{align*}
  &\ent{X_2|Y_2,X_1=x}
 \\{}={}
  &\sum_{y\in\dom{Y}} P_{x,\hat{x}} \log\frac{R_{x,y}}{P_{x,\hat{x}}}- \sum_{y\in\dom{Y}}\sum_{x'\in\preim{y}\setminus\{\hat{x}\}} P_{x,x'} \log\frac{P_{x,x'}}{R_{x,y}}
 \\\stackrel{(a)}{\le}{}
  &\sum_{y\in\dom{Y}} \left(R_{x,y}- P_{x,\hat{x}} \right)
   -\sum_{y\in\dom{Y}}\sum_{x'\in\preim{y}\setminus\{\hat{x}\}}
    \varepsilon \log\frac{\varepsilon}{R_{x,y}}
 \\\stackrel{(b)}{\le}{}
  &\sum_{y\in\dom{Y}} \left(R_{x,y}- P_{x,\hat{x}} \right)
  -\sum_{y\in\dom{Y}}\sum_{x'\in\preim{y}\setminus\{\hat{x}\}}
    \varepsilon \log\varepsilon
 \\\stackrel{(c)}{\le}{}
  &\sum_{y\in\dom{Y}}
   \left(\varepsilon-\varepsilon \log\varepsilon\right)
   \left(\card{\preim{y}}-1\right)
 \\={}
  & (N-M)\varepsilon \left(1- \log\varepsilon\right)
 \,,
\end{align*}
where $(a)$ is because $\log(1+x)\le x$, for $x'\neq\hat{x}$, $P_{x,x'}\le\varepsilon$ and $-p\log p$ increases on $[0,1/e]$, $(b)$ follows because $R_{x,y}\le 1$, and $(c)$ is due to~\eqref{eq:boundR}.

For the second inequality in~\eqref{eq:epserr}, because $\log(1+x)\le x$, we have $ N\varepsilon(1-\varepsilon)\le -N(1-\varepsilon)\log (1-\varepsilon)$. By assumption, $\varepsilon N<1$, whence $N\varepsilon(1-\varepsilon) \ge (N-M)\varepsilon$, for all $M\ge 1$. Thus,
\begin{align*}
 (N-M)\varepsilon \left(1- \log\varepsilon\right)
 ={}
  &(N-M)\varepsilon - (N-M)\varepsilon\log\varepsilon
 \\\le{}
  & N\varepsilon(1-\varepsilon) - N\varepsilon\log\varepsilon
 \\\le{}
  & -N(1-\varepsilon)\log (1-\varepsilon)- N\varepsilon\log\varepsilon
 \\={}
  &N\binent{\varepsilon}
 \,.\tag*{\qed}
\end{align*}

\subsection{Proof of Proposition~\ref{prop:optimal}}
\label{proof:optimal}

The set of all unrealizable length-$K$ sequences $\dom{X}^K\setminus\dom{S}_K$ is a clique in $\Ggraph_K$ and every vertex in this clique is connected to every vertex outside of it.
To see this fact, take $x\in\dom{X}^K$ such that $\Prob{\mblocked{X}_n=x}=0$.
Since this state can not be accessed, w.l.o.g. the $x$-th column of the corresponding adjacency matrix is zero.
This means that, for every $x'\in\dom{X}^K$, realizable or not, $\{x,x'\}\in E_K$.

Since $\dom{X}^K\setminus\dom{S}_K$ is a clique, and since every state in this clique is connected to an arbitrary $x\in\dom{S}_K$, also $\{x\}\cup(\dom{X}^K\setminus\dom{S}_K)$ is a clique.
A trivial clique partition thus consists of this clique and all the trivial single vertex cliques of vertices in $\dom{S}_K\setminus\{x\}$.
This clique partition has size $\card{\dom{S}_K}$. Since this clique partition may not be optimal, we get $M_K=\gamma(\Ggraph_K)\le\card{\dom{S}_K}$.

For the asymptotic result, note that $\lim_{K\to\infty} (\log M_K)/K$ cannot be smaller than $\entrate{\Xvec}$. But since $\entrate{\Xvec}=\log\lambda$ is achievable, we have $\lim_{K\to\infty} (\log M_K)/K\ge \log\lambda$. Furthermore, the number of realizable length-$K$ sequences of a Markov chain behaves like $\lambda^K$ as $K$ increases. Specifically, $\lim_{K\to\infty} (\log\card{\dom{S}_K})/K=\log\lambda$~\cite{Burda_MERW}. Together with $M_K\le\card{\dom{S}_K}$, this establishes~\eqref{eq:optimal}.\qed

\section{Zero-Error Source Coding of Stationary Processes}
\label{sec:discussion}

Based on the classic papers~\cite{Shannon_ZeroError} and~\cite{Witsenhausen_Confusability}, most results in zero-error information theory are based on memoryless channels and sources. While there exist extensions to channels with memory, see~\cite{Cohen_Fachini_Koerner__ZeroErrorCapacityOfBinaryChannelsWithMemory__IEEE_TIT_2016} and the references therein, to the best of the authors knowledge sources with memory have not been dealt with yet. We believe that applying zero-error information theory to Markov chains motivates such an extension. This section presents a first result.

Assume the source produces two jointly stationary random processes $\Xvec$ and $\Zvec$, and assume that the support of the marginal distribution is $\dom{S}_1:=\{(x,z)\in\dom{X}\times\dom{Z}{:}\ \Prob{X_n=x,Z_n=z}>0\}$. Furthermore, let $\dom{S}_K$ be the support of the joint distribution of $K$ samples, i.e., the joint distribution of $(X_1^K,Z_1^K)$. Clearly, $\dom{S}_K\subseteq\dom{S}_1^K$. We already mentioned that the $K$-fold co-normal product $\Ggraph_{(X,Z)}^{\vee K}$ of $\Ggraph_{(X,Z)}$ is the characteristic graph of the $K$-fold blocked source, assuming that the source $(X,Z)$ is iid~\cite{Korner_ZeroError}. We claim that independence is not necessary, but that $\dom{S}_K=\dom{S}_1^K$ suffices. As soon as $\dom{S}_K\subset\dom{S}_1^K$, the edge set of $\Ggraph_{\mblocked{(X,Z)}}$ may become a strict superset of the edge set of $\Ggraph_{(X,Z)}^{\vee K}$: Only deterministic dependence, where not all sequences $(x_1^K,z_1^K)$ are realizable, can reduce the required alphabet size as compared to the iid assumption.

If the receiver obtains the side information via a discrete, memoryless channel, the we get
\begin{prop}\label{prop:ZEprocess}
Let $\Xvec$ be a stationary stochastic process with support $\dom{S}_K$ of the distribution of $X_1^K=\mblocked{X}_1$ given as in Proposition~\ref{prop:optimal}, and let the side information $Z_1^K$ be given via a DMC $\Wvec$, i.e.,
\begin{equation}
 \Prob{X_1^K=x_1^K,Z_1^K=z_1^K}
 =\Prob{X_1^K=x_1^K} \prod_{i=1}^K W_{x_i,z_i}
 \,.
\end{equation}
Then, the characteristic graph $\Ggraph_{\mblocked{(X,Z)}}$ has edge set
\begin{equation}
 E_{\mblocked{(X,Z)}}
 = E_{(X,Z)}^{\vee K} \cup \{\{x,x'\}{:}\ x\in\dom{X}^K,x'\in\dom{X}^K\setminus\dom{S}_K\}
 \,.
\end{equation}
\end{prop}
Proposition~\ref{prop:ZEprocess} states that a deterministic temporal structure of the source can only decrease the clique partition number, making compression more efficient.
If $E_{\mblocked{(X,Z)}}=[\dom{X}^K]^2$, for some $K$, then no information needs to be transmitted because all information about $X_1^K$ is already contained in the side information $Z_1^K$. We believe that this analysis can be extended to more general side information structures and to variable-length zero-error source codes as in~\cite{Koulgi_ZESC,Alon_GraphEntropies}.

\begin{proof}
By Definition~\ref{def:chargraph}, $\{x,x'\}\in E_{\mblocked{(X,Z)}}$, iff, for all $z\in\dom{Z}^K$,
\begin{equation}\label{eq:condInProof}
 \Prob{X_1^K=x,Z_1^K=z}\Prob{X_1^K=x',Z_1^K=z}=0
 \,.
\end{equation}
With $x_i$ the $i$-th coordinate of $x$, we write
\begin{equation}
 \Prob{X_1^K=x,Z_1^K= z} = \Prob{X_1^K=x} \prod_{i=1}^K W_{x_i,z_i}
\end{equation}
and see that~\eqref{eq:condInProof} holds, iff at least one of the following conditions holds:
\begin{subequations}\label{eq:conditions}
\begin{align}
 \label{eq:imp1}
 \Prob{X_1^K=x} &= 0
 \,,
 \\
 \label{eq:imp2}
 \Prob{X_1^K=x'} &= 0
 \,,
 \\
 \label{eq:ORnot}
 \prod_{j=1}^K\prod_{i=1}^K W_{x_i,z_i}W_{x'_j,z_j} &= 0
 \,.
\end{align}
\end{subequations}
Equation~\eqref{eq:imp1} (and, similarly, equation~\eqref{eq:imp2}) imply that if a sequence $x$ is not realizable, then~\eqref{eq:condInProof} holds for all $x'\in\dom{X}^K$. Hence, in $\Ggraph_{\mblocked{(X,Z)}}$, each unrealizable state $x$ is connected to every other state. With $\dom{S}_K$ being the set of realizable sequences, we get $\{\{x,x'\}{:}\ x\in\dom{X}^K,x'\in\dom{X}^K\setminus\dom{S}_K\}\subseteq E_{\mblocked{(X,Z)}}$.

We may assume w.l.o.g. that $\dom{S}_1=\dom{X}$, i.e., that all states are realizable. Then, since the $K$-fold co-normal product $\Ggraph_{(X,Z)}^{\vee K}$ of $\Ggraph_{(X,Z)}$ is the characteristic graph of the source emitting $(X,Z)$ iid, we have $\{x,x'\}\in E_{(X,Z)}^{\vee K}$, iff, for all $z\in\dom{Z}^K$,
\begin{equation}
 \prod_{j=1}^K\prod_{i=1}^K \mu_{x_i}\mu_{x'_j}W_{x_i,z_i}W_{x'_j,z_j}=0
 \,.
\end{equation}
Since we assume that $\muvec>\mathbf{0}$, this is equivalent to~\eqref{eq:ORnot}. Hence, also $E_{(X,Z)}^{\vee K}\subseteq E_{\mblocked{(X,Z)}}$. This covers all cases of~\eqref{eq:conditions}.
\end{proof}

\Acknowledgements{}
\Bibliography{}
\end{document}